\documentstyle[prd,aps]{revtex}
\input psfig.tex

\begin{document}
\draft
\title{Cosmic Birefringence within the Framework of Heterotic String Theory} 
\author{Prasanta Das\footnote{Electronic address: {
pdas@iitk.ac.in}}, Pankaj Jain\footnote{Electronic address: 
{pkjain@iitk.ac.in}}, and Sudipta Mukherji\footnote{Electronic 
address: {mukherji@iitk.ac.in}}}
\address{Department of Physics, 
\\Indian Institute of Technology, Kanpur 208 016, India}

\date{November 2000}
\maketitle
\begin{abstract}
{ 
Low energy string theory predicts the existence of an axion field which can 
lead to cosmic birefringence. We solve the electromagnetic wave
equations in the presence of such an axion and a dilaton field
in order to determine their effect on
the polarization of light. We find that the presence of dilaton
field leads to a nontrivial modification of the final result. We comment
on the possibility of discovering such an effect by observations of
radio wave polarizations from distant radio galaxies and quasars. We have
also determined the limits on the string theory parameters that are imposed by
the current radio polarization data.  
}
\end{abstract}

\begin{flushleft}
~~~~~~~~~~~~~~~~~~Pacs no. 14.80.Mz,98.80.Cq,11.25.-w,98.54.- h 
\end{flushleft}

\maketitle

\section{Introduction}
There have been several studies which explore the possible
existence of cosmic birefringence
\cite{bire}. 
In recent work Kar et al \cite{Kar} and Majumdar and Sengupta \cite{Partha} 
has pointed out that the Kalb-Ramond field, which arises in supergravity
theories, makes the space time birefringent. The authors 
solve the electromagnetic
wave equations in the presence of such a field
and predict that this field will lead
to a redshift dependent rotation of polarization whose details
depend on whether the universe is radiation or matter dominated.  
Earlier in Ref. \cite{CFJ}
Carroll et al analyzed the polarizations of radio waves from distance
galaxies and quasars as a function of redshifts and found no
statistically significant effect. 
The observable of interest in that study was the angle $\beta=\chi-\psi$,
where $\psi$ is the orientation angle of
the axis of the radio galaxy and $\chi$ the observed polarization angle
after the effect of Faraday rotation is taken out of the data.
This can be done by making a straight line fit to the polarization 
angle $\theta(\lambda^2)$ which can be expressed as
\begin{equation}
\theta(\lambda^2) = RM\ \lambda^2 + \chi
\end{equation}
where $\lambda$ is the wavelength, $RM$ and $\chi$ are the slope and
the intercept of the fit. $RM$ stands for rotation measure and
depends on the plasma density and the parallel component of the background 
magnetic field along the direction of propagation of the wave. 
Later Ref. \cite{NR} claimed the existence of a 
large scale anisotropy in the same data. This claim was, however,
questioned by many authors \cite{EB,CF97,LFW}. 

The possibility of local anisotropy, independent of redshift,
was explored in Ref. \cite{jain98} and earlier in \cite{Birch,KY}. 
These authors found a statistically significant
effect. The original claim made by Birch \cite{Birch} in 1982 was dismissed by
Bietenholz and Kronberg \cite{BK} in 1984 who compiled a larger data
set which they claimed did not show any effect. However Jain and Ralston
\cite{jain98} argued that Ref. \cite{BK} did not pay careful attention to the
parity of the correlation ansatz \cite{RJ}. When the transformation
properties of statistical distribution are taken into account the data
does show a significant effect which is further enhanced by choosing 
an appropriate statistic as well as by putting some cuts on the data.

Inspite of the progress in our understanding in string theory in the
last two decades, we are yet to have a string theory prediction that can
be tested in the experiments. One of the main reason for such an
unsatisfactory situation is that the energy scale at which stringy effects
become significant is beyond the reach of present day experiments. It is
thus believed that the most likely area for a confrontation between string
theory and experiments is through extracting astrophysical or
cosmological consequences of string theory which might be measurable
today. With this kind of motivation in mind, we study in this paper the
prediction of string theory for cosmic birefringence. However, as our
knowledge of string theory is mostly perturbative in nature, we face
several problems in carrying out such a study. These are mainly due to
the presence of various massless moduli that are present in the 
low energy limit of string theory. We will point those out as we proceed.

Four dimensional heterotic string theory, in its low
energy limit, contains  moduli coming from compactification, gauge
fields, dilaton and an axion. In this theory, coupling among various
fields are unique and
they are all controlled by the expectation value of dilaton. Given such a
framework, we analyse the propagation of electromagnetic waves
in curved metric background. Due to the
presence of dilaton, as we will argue, medium turns into a conducting
medium where the time variation of dilaton plays the role of effective
conductivity. On the other hand, the axion  rotates the plane of
polarization of
the electromagnetic wave causing non-trivial birefringence. 
However, notice that the dilaton (and other moduli) is a massless scalar
in string theory. An important issue to resolve in string cosmology is to
explain the absence of massless dilaton at the present time. One expects
that a potential for dilaton will be generated in string theory through
some non-perturbative effects and the dilaton will sit at the minimum of
the potential picking up a mass. The mechanism of how precisely this should
happen is still unknown in string theory. This, in turn, creates the most
severe problem in comparing observational data with  string
theory. In the rest of the paper, we continue to treat dilaton as a
massless scalar and find string prediction for cosmic birefringence. We
hope that some of our results will be helpful when dilaton decoupling in
string theory is properly understood.
In general, we
find that the relevant equations are hard to solve exactly. However,
it is well-known that heterotic string theory in four dimensions is 
self dual. The duality group is parametrized by $SL(2,R)$ matrices. Using
this
property, in the next section, we  generate some non-trivial solutions
for various fields starting from some simple seed configuration. This, in
turn,
allows us to calculate birefringence 
 of electromagnetic waves, propagating in radiation or matter
dominated universe, in a fairly straightforward manner. 
However, since the solution generating technique does
not saturate all possible allowed configurations, in section III, we 
solve the electromagnetic equations within the WKB approximation. 
We then discuss our results in section IV.

\section{Birefringence from Heterotic String} 

In the low energy limit, the heterotic string compactified on six-torus
is represented by the following action \cite{ashokesen}:
\begin{eqnarray}
S &=& \int d^4x {\sqrt{-G}}e^{-\phi}\big[ R_{G} +
G^{\mu\nu}\partial_\mu\phi\partial_\nu\phi - {1\over
{12}}H_{\mu\nu\rho}H^{\mu\nu\rho} ~\nonumber \\
&-& G^{\mu\rho}G^{\nu\lambda}F_{\mu\nu}^{(a)}(LML)_{ab}
{F^{(b)}}_{\rho\lambda} + {1\over 8} G^{\mu\nu} {\rm Tr}(\partial_\mu M
L \partial_\nu M L)\big] \label{het}
\end{eqnarray}
where the gauge fields and the anti-symmetric three rank tensor field
are defined as
\begin{eqnarray}
F_{\mu\nu}^{(a)} &=& \partial_\mu A_\nu^{(a)} - 
\partial_\nu A_\mu^{(a)}, \nonumber \\
H_{\mu\nu\rho} &=& (\partial_\mu B_{\nu\rho} + 2 A_\mu^{(a)}L_{ab}
F_{\nu\rho}^{(b)}) + {\rm cyclic ~permutations ~of } ~\mu, \nu, \rho. 
\label{hfdef}
\end{eqnarray}
The superscript $a$ on $F_{\mu\nu}$ runs from $1$ to $28$ denoting
the $28$ gauge fields that appear in the four dimensional heterotic 
string action. Out of these $28$ gauge fields, $16$ are originally
present in ten dimensions. The rest $12$ appear due to torus
compactification. Among other terms in (\ref{het}), $R_G$ is the 
scalar curvature associated with the metric $G_{\mu\nu}$ and $\phi$
is the dilaton. The $28 \times 28$ dimensional matrix $M$ encodes
all the scalars that appear due to compactification. \footnote{
Note that in  (\ref{het}), the first two terms in the action is 
like Brans-Dicke (BD) theory with the BD parameter $\omega = -1$. The
radar-echo delay experiments \cite{rs}
in solar system have set the limit on
$\omega > 500$. So in a sense, the theory as it stands is ruled out by
this experiment. However, let us note that the string action in general
has many moduli field (such as matrix $M$ here). If these scalars happen
to have flat direction along the dilaton, the effective $\omega$ might
change from its value $-1$.}
 In what follows, the
detail of their structure will not be important. The matrix $L$ in 
(\ref{het}) is also $28 \times 28$ dimensional and is given by
\begin{equation}
L = \pmatrix{0&{I_6}&0\cr{I_6}&0&0\cr0&0&{-I_{16}}\cr} \nonumber
\end{equation}
where $I_n$ denotes $n \times n$ identity matrix.

In the following, however, we will work with a truncated version
of the effective action. We will turn off all the scalars that 
appear due to compactification. This essentially amounts to setting
the matrix $M$ to identity. Furthermore, we will set all the gauge 
fields to zero except one. We will take this one to be present 
in the original ten dimensional heterotic action. We will see that
this choice of trunctation makes the dynamics tractable and at the
same time keeps the essential physics intact. In this approximation,
the action, in the Einstein frame, reduces to
%
\begin{equation}
S = \int d^4x {\sqrt{-g}}[R_g - {1\over 2} g^{\mu\nu}\partial_\mu\phi
\partial_\nu\phi - {1\over {12}}e^{-2\phi}H_{\mu\nu\rho}H^{\mu\nu\rho} 
- {e^{-\phi}}F_{\mu\nu}F^{\mu\nu}] \label{einhet}
\end{equation}
The equation of motion for the $B_{\mu\nu}$ 
following from the above action is 
\begin{equation}
D_\mu(e^{-2\phi}H^{\mu\nu\rho}) = 0.
\end{equation}
This, in turn, allows us to introduce a pseudoscalar field $\Psi$, known as
axion, through the relation
\begin{equation}
H^{\mu\nu\rho} = - {1\over {\sqrt{-g}}}e^{2\phi}
\epsilon^{\mu\nu\rho\sigma} \partial_\sigma \Psi.
\end{equation}
With this definition of $\Psi$, the rest of the equations
of motion are the following
\begin{eqnarray}
&&R_{\mu\nu} = {1\over 2} e^{2\phi}\partial_\mu \Psi \partial_\nu \Psi
     + 2 e^{-\phi}F_{\mu\rho}F_{\nu}^{\rho} - {1\over 2} g_{\mu\nu}
     e^{-\phi} F_{\rho\sigma}F^{\rho\sigma} + {1\over 2} \partial_\mu\phi 
     \partial_\nu \phi, \label{curv} \\
&&D_\mu(e^{-\phi} F^{\mu\nu} + \Psi {\tilde F}^{\mu\nu}) = 0,
\label{gfield}\\
&&e^{2\phi}D^\mu D_\mu \Psi + 2e^{2\phi}\partial_\mu \phi
\partial^\mu \Psi - {\tilde F}_{\mu\nu}F^{\mu\nu} = 0,
\label{sai}\\
&& e^{2\phi}D^\mu D_\mu (e^{-\phi}) + e^{3\phi} \partial_\mu \Psi
\partial^\mu \Psi - e^{\phi}\partial_\mu \phi \partial^\mu \phi
- F_{\mu\nu}F^{\mu\nu} = 0.\label{phieq}
\end{eqnarray}
Here 
\begin{equation}
{\tilde F}^{\mu\nu} = {1\over {2
{\sqrt{-g}}}}e^{\mu\nu\rho\sigma}F_{\rho\sigma},\label{dualg}
\end{equation}
which satisfies Bianchi identity
\begin{equation}
D_\mu {\tilde F}^{\mu\nu} = 0.\label{bianc}
\end{equation}

It is easy to check that the set of equations (\ref{curv})-(\ref{phieq})
and (\ref{bianc}) are invariant under the following set of $\rm{SL}(2,R)$
transformations \cite{ashokesen}

\begin{equation}
\lambda \rightarrow \lambda^\prime = {{ a\lambda + b}\over {c\lambda +
d}}, ~~F_{\mu\nu} \rightarrow F^\prime_{\mu\nu} = (c \Psi + d)F_{\mu\nu} -
c e^{-\phi}{\tilde
F}_{\mu\nu}, ~~g_{\mu\nu} \rightarrow g_{\mu\nu},\label{dtrans}
\end{equation}
where $\lambda = \Psi + i e^{-\phi}$ and $a,~b,~c,~d$ are real parameters
with $ad - bc =1$. 

Here, in what follows, we will restrict ourselves to the case where
dilaton, axion and the metric are  only function of time. Our startegy
will be to solve all the equations of motion in a {\it fixed} metric
background. First, we take the metric as 
\begin{equation}
ds^2 = R^2{(\eta)}( - d\eta^2 + dx^2 + dy^2 + dz^2 ).
\label{met}
\end{equation}
The gauge field strength $F_{\mu\nu}$ in this background 
can be defined as \cite{CF91}
\begin{equation}
F_{\mu\nu} = R^2(\eta)\pmatrix{0&{-E_x}&{-E_y}&{-E_z}\cr
{E_x}&0&{B_z}&{-B_y}\cr
{E_y}&{-B_z}&0&{B_x}\cr
{E_z}&{B_y}&{-B_x}&0\cr}
\end{equation}
With this definition, equation (\ref{gfield})  reduces to
\begin{equation}
\nabla \cdot {\bf E} = 0,~~\partial_\eta(e^{-\phi}{\bf E} R^2)
- \nabla \times (e^{-\phi}{\bf B} R^2) = - 2 \partial_\eta\Psi {\bf
B}R^2
\end{equation}
Similarly the scalar equations are
\begin{eqnarray}
&&\partial_\eta \partial^\eta \Psi -{2\over {R^3}} \partial_\eta R 
\partial_\eta \Psi - {2\over R^2} \partial_\eta \phi \partial_\eta \Psi
+ 4 R^2 e^{-2\phi}{\bf E}\cdot{\bf B} = 0, \label{one}\\
&&\partial_\eta \partial^\eta \phi - {2\over {R^3}} \partial_\eta R
\partial_\eta \phi + {e^{2\phi}\over {R^2}}(\partial_\eta \Psi)^2
 - 2 R^2 e^{-\phi}(E^2 - B^2)
= 0. \label{two}
\end{eqnarray}
On the otherhand, (\ref{bianc}) reduces to
\begin{equation} 
\nabla \cdot {\bf B} = 0,~~
\partial_\eta (R^2{\bf B}) + \nabla \times (R^2{\bf E)} = 0.\label{three}
\end{equation}

\subsection{\bf{Analyzing Equations in flat background}}

Before we go on to analyse more realistic scenario, it is worthwile to
study (\ref{one}) - (\ref{three}) in {\it flat} metric background 
for which $R(\eta) = 1$. Beside  being completely tractable, this will
indeed give us an idea of dilatonic and axionic influence on the
propagation of electromagnetic waves.
In this very special context, equations (\ref{one})-(\ref{three}) reduce
to the following forms:
\begin{eqnarray}
&&\partial_\eta^2 \Psi + 2 \partial_\eta\phi \partial_\eta \Psi = 0,
\label{four} \\
&&\partial_\eta^2\phi - e^{2\phi} (\partial_\eta\Psi)^2 + 2e^{-\phi}(
{\bf E}^2 - {\bf B}^2) = 0,\label{five}\\
&& \partial_\eta(e^{-\phi}\partial_\eta{\bf B}) - \nabla^2(e^{-\phi}
{\bf B}) = 2 \partial_\eta \Psi  \nabla\times{\bf B}.
\label{six}
\end{eqnarray}
It is easy to check that when $\phi$ is turned of, the above equations 
reduce to the case analyzed in \cite{Kar}. The very presence of dilaton
$\phi$ effects the dynamics in a very crucial way. In order to understand
the behaviour of the solutions, we first set the axion field $\Psi$ to
zero. Then equation (\ref{six}) reduces to 
\begin{equation} 
\partial_\eta^2 {\bf B} - \partial_\eta\phi  \partial_\eta{\bf B} -
\nabla^2{\bf B} =
0.\label{magfield}
\end{equation}
Such an equation for  magnetic field is quite well known from the studies 
of electromagnetic wave propagation in conducting medium. 
As we can see from (\ref{magfield}), the time variation of dilaton plays
the role of effective conductivity of the medium.  Such
a damping factor, as usual, reduces the amplitude of the magnetic field
$\bf B$. 

The set of equations (\ref{three})-(\ref{six}) are in general very hard to
solve directly. Though it would be desirable to have such solutions, we
would however use the property (\ref{dtrans}) of heterotic string theory
to generate new solutions from some simple ones. To this end, we notice
that if we set the dilaton field to zero (using sift symmetry of dilaton
in heterotic string theory, we can always bring any constant dilaton field
to zero), equations
(\ref{three})-(\ref{six}) possess a set of solutions \cite{Kar}
\begin{eqnarray}
\Psi = \int h dt =  ht, ~~~B_{\pm}(t,z) 
= {b_\pm} e^{i k z} e^{i \omega_\pm t},~{\rm with}~{ E}_{\pm}  
= \pm \frac{1}{k}\frac{d}{d\eta}(B_{\pm})= \pm{i \omega_\pm\over k}{B}_{\pm},
\label{seed}
\end{eqnarray}
where $h$ is an integration constant and $\omega_\pm = {\sqrt{k^2 \pm
2hk}}$. In the above equations, we have defined ${\bf B} (t,z) = 
{\bf b}(t)e^{ikz}$ with $z$ being the direction of propagation.
Furthermore $b_\pm$ are defined as the circular polarization states,
namely $b_\pm = b_x \pm i b_y$.
Now writing $B_\pm = b_\pm e^{ikz}e^{i\varphi_\pm}$ and birefringence as 
$\delta \varphi = {1\over 2} (\varphi_+ - \varphi_-)$, we get \cite{Kar}
\begin{eqnarray}
\delta \varphi = h (t_2 - t_1), ~{\rm 
for}~k >>h. 
\end{eqnarray}
Here, for flat metric, $\eta$ is same as cosmic time $t$. Thus in the 
above equation, $t_1$ and $t_2$ are the earlier and present time. 

As discussed earlier in (\ref{dtrans}), heterotic string theory in four
dimension has self-duality symmetry. As a consequence, equations of motion
for various fields are invariant under $SL(2,R)$ transformation. 
These transformations have the property to generate non-trivial field 
configurations from the known ones. A generic solution is thus 
parametrized by three independent parameters of $SL(2,R)$ group. Here,
we  start with  the field configuration given in (\ref{seed})
where dilaton field is set to zero. Next, by the use of $SL(2,R)$
symmetry, we  generate configurations 
with nontrivial dilaton and various other fields. Strictly following the
transformation rules given in (\ref{dtrans}), and denoting
the new configurations by $(\Psi^\prime, \phi^\prime, {\bf B}^\prime,
{\bf E}^\prime )$, we, thus, get
\begin{equation}
\Psi^\prime = {{ ac \Psi^2 + (bc + ad)\Psi + (ac+bd)}\over
{(c\Psi + d)^2 + c^2}}, ~~e^{-\phi^\prime} = {1\over 
{(c \Psi  + d)^2 + c^2}}.\label{newaxdil}
\end{equation}
where $\Psi$ is given in (\ref{seed}).
As expected the solutions are parametrized by three independent
${\rm SL}(2,R)$ parameters (taken here as) $b, c, d$.
The other parameter $a$ is determined from the relation $ad - bc =1$.
Furthermore, from
the transformation property of $F_{\mu\nu}$ under $SL(2,R)$, we get
\begin{eqnarray}
&&{ B}_\pm^\prime = [(c \Psi+ d) \mp i c]b_\pm e^{ikz}
e^{i\omega_\pm t}
 \nonumber\\
&&{ E}_\pm^\prime = \pm{i\omega_\pm \over k}[(c \Psi + d) \mp i c]b_\pm
e^{ikz} e^{i\omega_\pm t}, ~{\rm for}~k>>h 
\end{eqnarray}
Thus the birefringence is then given by
\begin{equation}
\delta \varphi^{\prime} = h(t_2 - t_1) 
 - \frac{1}{2}\left[\tan^{-1}\left(\frac{2(c\Psi + d)c}{(c \Psi + d)^2 - c^2}
\right)\right]_{t_1}^{t_2}\ \ \ \  ~{\rm for}~k >>h.
\end{equation}

\subsection{Analyzing the equations in curved space}

Here, we will study (\ref{gfield})-(\ref{bianc}) for the general
metric given in (\ref{met}). We will follow the same strategy 
as the previous subsection in order to find solutions of equations of
motion. Keeping this in mind, let us first consider
(\ref{gfield})-(\ref{bianc}) when the dilaton is set to zero. In that
case, defining $F_\pm = R^2 (\eta)B_\pm$, 
we get an equation
for magnetic field as follows \cite{Kar}:
\begin{eqnarray}
&&{d^2F_\pm\over {d \eta^2}} + (k^2  \pm {2 h k\over 
{R^2(\eta)}})F_\pm 
= 0,\nonumber\\ 
&&\Psi = h \int {1\over {R^2(\eta)}} d\eta,
\label{xyz}
\end{eqnarray}
with $h$ being a constant. If $R(\eta)$ increases fast enough with $\eta$
(which is indeed the case for radiation and matter dominated universe), a
WKB type analysis will give us a very close to exact result for the above 
equation. Carrying out such a computation we get 
\begin{equation}
F_\pm = f_\pm e^{i k \eta} e^{i k z} 
e^{ \left[ \pm i \int {{h d\eta}\over {R^2(\eta)}} 
\right]} 
= f_\pm e^{i k \eta} e^{i k z} e^{i S_{\pm}},
\end{equation}
where we have $S_{\pm} = \pm \int {{h d\eta}\over R^2(\eta)} $.
Like $B_{\pm}$ as in eqn.(24), one can also rewrite the eqn.(30) in the following 
way,

\begin{eqnarray}
F_{\pm}= f_{\pm} e^{ikz} e^{i\omega_{\pm}^{R} \eta }
\end{eqnarray}
where, $\omega_{\pm}^{R} = \sqrt{(k^2 \pm \frac{2 h k}{R^2(\eta)})}$.\\

Now we have,
\begin{equation}
B_\pm = {1\over {R^2(\eta)}} F_{\pm}  = 
{1\over {R^2(\eta)}} f_\pm e^{i k \eta}e^{i k z}   e^{ \pm i \int 
{{h d\eta}\over {R^2(\eta)}}}, 
~~ {\rm and} ~~E_{\pm} = \pm {1\over {k R^2(\eta)}} {{d\over {d \eta}}(
B_{\pm} R^2(\eta))}\label{bpm}
\end{equation}
From the above experssions for $B_\pm$, we see that the birefringence due
to the presence of axion (dilaton is set to zero) is given by,
\begin{equation}
\delta \varphi = \frac{1}{2}(S_{+} - S_{-} ) = \int_{\eta_1}^{\eta_2}\frac{h d\eta}{R^2}  
=  \int_{t_1}^{t_2}\frac{h dt}{R^3}  
\end{equation}
where we have converted the conformal time $(\eta)$ into cosmic time $t$
and $t_1$ and $t_2$ are the earlier time and present time respectively. 

As before, starting from (\ref{xyz}) and (\ref{bpm}) 
and using the freedom of $SL(2,R)$ symmetry of
equations of motion, we generate here large class of configurations, where 
axion and dilaton are given by
\begin{equation}
\Psi^\prime = {{ac \Psi^2 + (bc + ad)\Psi + (ac+bd)}\over
{(c\Psi + d)^2 + c^2}}, ~~e^{-\phi^\prime} = {1\over
{(c \Psi  + d)^2 + c^2}},\label{naxdil}
\end{equation}  
where $\Psi$ is given in (\ref{xyz}). On the other hand,
from the transformation property of $F_{{\mu}{\nu}}$ under $SL(2,R)$, we
get
\begin{eqnarray}
B^{\prime}_{\pm} = \left[(c \Psi + d)
\mp i c (1 \pm {h \over {k R^2(\eta)}})\right]
B_{\pm}.
\label{bbb}
\end{eqnarray}
Consequently, the expression of birefringence  is given by
\begin{eqnarray}
\delta \varphi^\prime = - \frac{1}{2} \tan^{-1}\left[\frac{2(c \Psi + d) c }
{(c\Psi + d)^2 - c^2(1 - \frac{h^2}{k^2 R^4})}\right]
+  \int_{\eta_1}^{\eta_2} \frac{h d\eta}{R^2}
\end{eqnarray}  

After having  $\delta\varphi^{\prime}$ for general metric, let
us concentrate on two particularly interesting cases; first radiation
dominated
case and then matter dominated case.

\vspace*{0.15in}

\noindent{\bf B1. Radiation dominated Universe}: For radiation
dominated 
universe, $R$ is given
by $R(t) = R_{0R}t^{1/2}$, where $R_{0R}$ is a dimensionful constant.
Instead of writing various fields explicitly in this particular case, 
we give the expression of birefringence that follows from (\ref{bbb}).
It is given by \footnote{The relation that we used between $R(t)$ and $z$
can be described as
follows,
\begin{eqnarray}
\frac{R(t_2)}{R(t_1)} = 1 + z = \frac{(t_2)^{\alpha}}{(t_1)^{\alpha}}
\end{eqnarray}
where for Radiation dominated phase $\alpha = \frac{1}{2} $ and for matter
dominated phase $\alpha = \frac{2}{3}$}:
\begin{eqnarray}
\delta \varphi^\prime = - \frac{1}{2} \tan^{-1}\left[\frac{2c(c \Psi + d)}
{(c\Psi + d)^2 - c^2(1 - \frac{ h^2}{ k^2 R_{0R}^4 t_2^2})}\right]
-  \frac{h}{R_{0R}^3 t_2}\left[1 - (1 + z)^{2}\right]\ .
\end{eqnarray}

\vspace*{.15in}

\noindent {\bf B2. Matter dominated Universe}: For matter dominated phase,
$R$ is given by $R(t) = R_{0M}t^{2/3}$. In the same way as before, one can
show that the birefringence is now given by:
\begin{eqnarray}
\delta \varphi^\prime = - \frac{1}{2} \tan^{-1}\left[\frac{2c(c \Psi + d) }
{(c\Psi + d)^2 - c^2(1 - \frac{h^2}{ k^2 R_{0M}^4 (t_2)^{8/3}})}\right]
-  \frac{h}{R_{0M}^3 t_2}\left[1 - (1 + z)^{3/2}\right],
\end{eqnarray}

where, in both cases, the second term corresponds to the birefringence 
corresponding to zero dilaton field $\phi$.
However, in both  the  cases, we see that the expression of
$\delta\varphi^\prime$ depends on redshift $z$ in a crucial way.
Furthermore, though very weak, $\delta\varphi^\prime$ has a dependence on
the frequency $k$ of the electromagnetic wave. 

\section{Approximate Solutions including the Dilaton Field}

So far we have concentrated on solving the equations by first 
setting the dilaton field equal to zero and then using the $SL(2,R)$
transformation to generate solutions with nonzero dilaton field. 
However this procedure does not guarantee that all possible solutions
to the equations are generated. We next solve the field equations
directly within the WKB approximation. We first write the electromagnetic
field equation in terms of the variable $S_\pm$. Since $S_\pm$ is
assumed to be a slowly varying function we drop its second derivatives
and terms proportional to $[S'_\pm]^2$. With these approximations the
field equations reduce to
\begin{eqnarray}
{\partial S_\pm^R\over \partial t} &=& \mp {\partial \psi\over\partial t}
e^\phi\\
{\partial \psi\over \partial t} &=& {h\over R^3} e^{- 2\phi} \\
{\partial^2\phi\over\partial t^2} & = & - {3\over R}{\partial R\over \partial t}
{\partial\phi\over \partial t}  + {h^2\over R^6} e^{-2\phi}\label{dilaton_eqn}\\
S^I_\pm &=& -\phi/2
\end{eqnarray} 
where $S_\pm^R$ and $S_\pm^I$ are the real and imaginary parts of the 
function $S_\pm$. We also drop the term proportional to $h^2$ in the
dilaton field equation (\ref{dilaton_eqn}) 
since it is higher order in $h$ and hence 
expected to give negligible contribution.
We then find that Eq. \ref{dilaton_eqn} can be solved to give
$$ \phi = B + A/t$$
where $A$ and $B$ are constants. Furthermore, 
$S_\pm^R$ are given by, 
\begin{equation}
S^R_\pm = \mp {h e^{-B}\over AR_{0M}^3}e^{-A/t_2}\left[1-\exp\left(-{A\over t_2}
\left((1+z)^{3/2} -1\right)\right)\right]
\label{approximate}
\end{equation}
The rotation in polarization is given by $\delta \phi = (S_{+} - S_{-})/2$.
The functional form of the rotation angle is plotted in Fig. 1 as
a function of redshift for some arbitrary choice of the parameters
$\zeta=-e^{-A/t_2}h e^{-B}/AR_{0M}^3$ and $\xi = A/t_2$.
Hence we find that the dilaton field leads to a nontrivial change in
the redshift dependence of the rotation in polarization. Whereas in the
absence of dilaton we predict a steady increase of the polarization
rotation angle with redshift, in the present case the rotation turns on 
only at large redshifts.  

\begin{figure}
\psfig{file=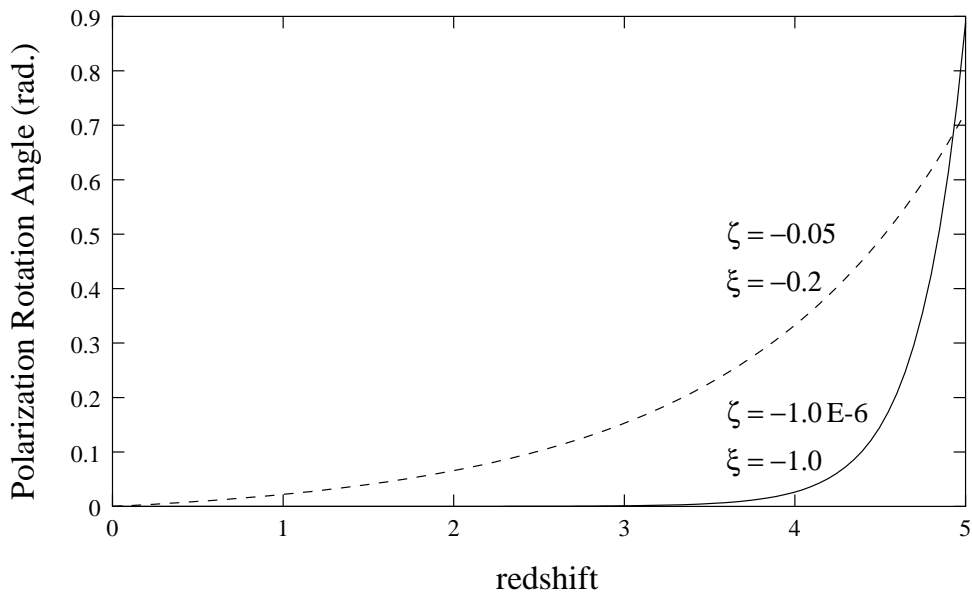}
\bigskip
\caption{The redshift dependence of the polarization angle for some
representative values of the parameters $\zeta$ and $\xi$}
\end{figure}

We next consider the polarization data from distant quasars and radio 
galaxies in order to determine if there exists a signal of the type
predicted by our analysis.
We considered all the sources that were used in the
statistical analysis in Ref. \cite{jain98} for which the information
about redshift was available. The NASA extragalactic data base (NED)
was used to update the information about redshifts. We compiled a
total of 231 sources in this 
manner\footnote{The entire data set is available on the website 
$home.iitk.ac.in/\tilde{\   } pkjain$}. 
We also examined the data after removing 
all sources which lie within $\pm 30^o$ of the galactic plane. This cut
is useful since it removes many sources with very large rotation
measures and hence reduces the possibility of bias in data. The
bias in rotation measure can
arise due to the $n\pi$ ambiquity in the measurement of linear polarization.
After imposing this cut on the data we are left with a total of
160 sources.
We use the maximum likelihood analysis in order
to determine the presence of correlation in data. 
For the null hypothesis we use the 
von Mises (vM) distribution which serves as a prototype for 
the statistical fluctuations for circular data. It is given by \cite{Fisher},
\begin{equation}
f_{\rm vM}(\beta) = {\exp \left[ \kappa\cos2(\beta-\overline \beta)
\right] \over \pi I_0(\kappa)}
\label{von}
\end{equation} 
where $\kappa$ is a parameter which measures the concentration of the
population, $\beta$ is the angle between the axis of the radio galaxy
and the observed polarization after taking out the effect of Faraday
rotation and $\overline \beta$ is the mean angle. The factor
2 has been inserted since the polarization angle is ambiguous by $n\pi$
and not $2n\pi$ \cite{RJ}. The distribution peaks very
close to $\overline\beta=\pi/2$ and therefore it is reasonable to
eliminate this parameter by setting it exactly equal to $\pi/2$ \cite{jain98}.
We are therefore left with only one free parameter and the fit to
the entire data set leads to $\kappa = 0.67$ and after the galactic
cut we find $\kappa = 0.71$.

\begin{figure} [t,b] \hbox{\hspace{6em}
\hbox{\psfig{figure=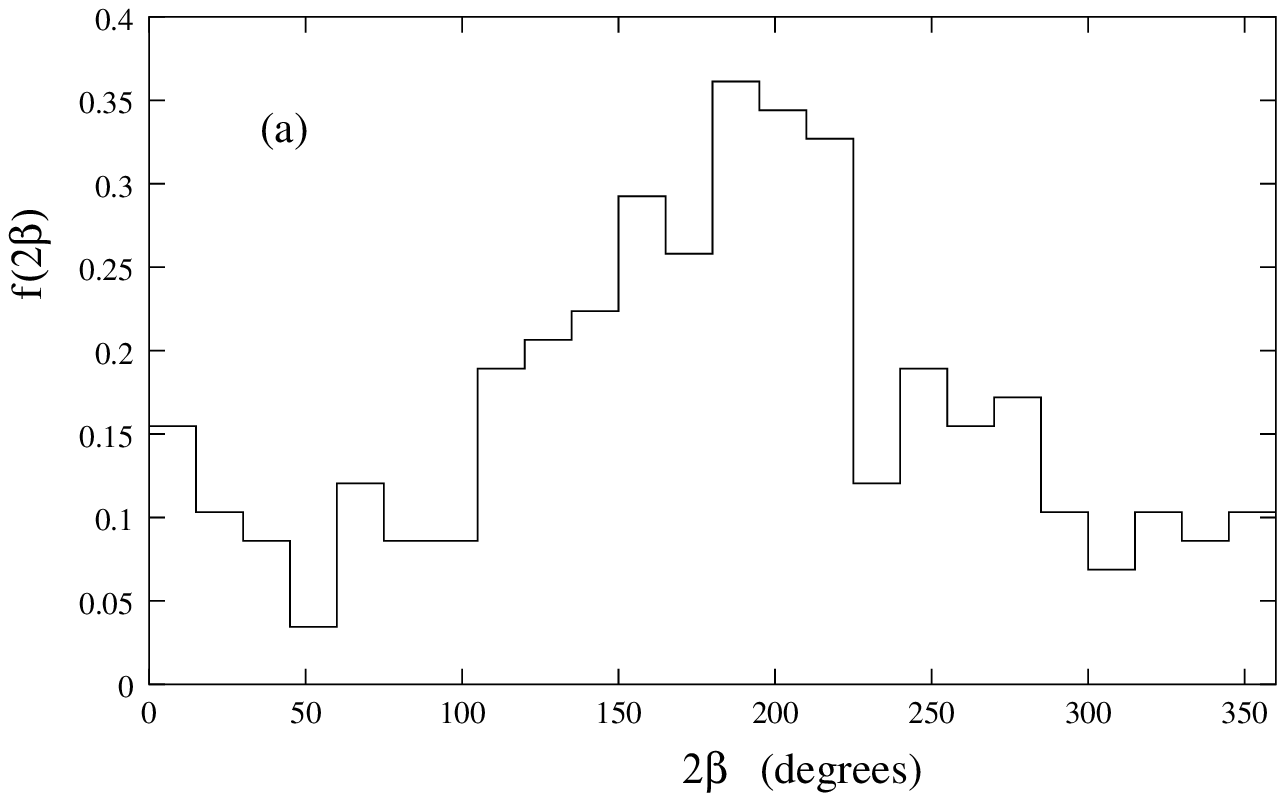,height=6cm}}}
\bigskip
\hbox{\hspace{6em}
\hbox{\psfig{figure=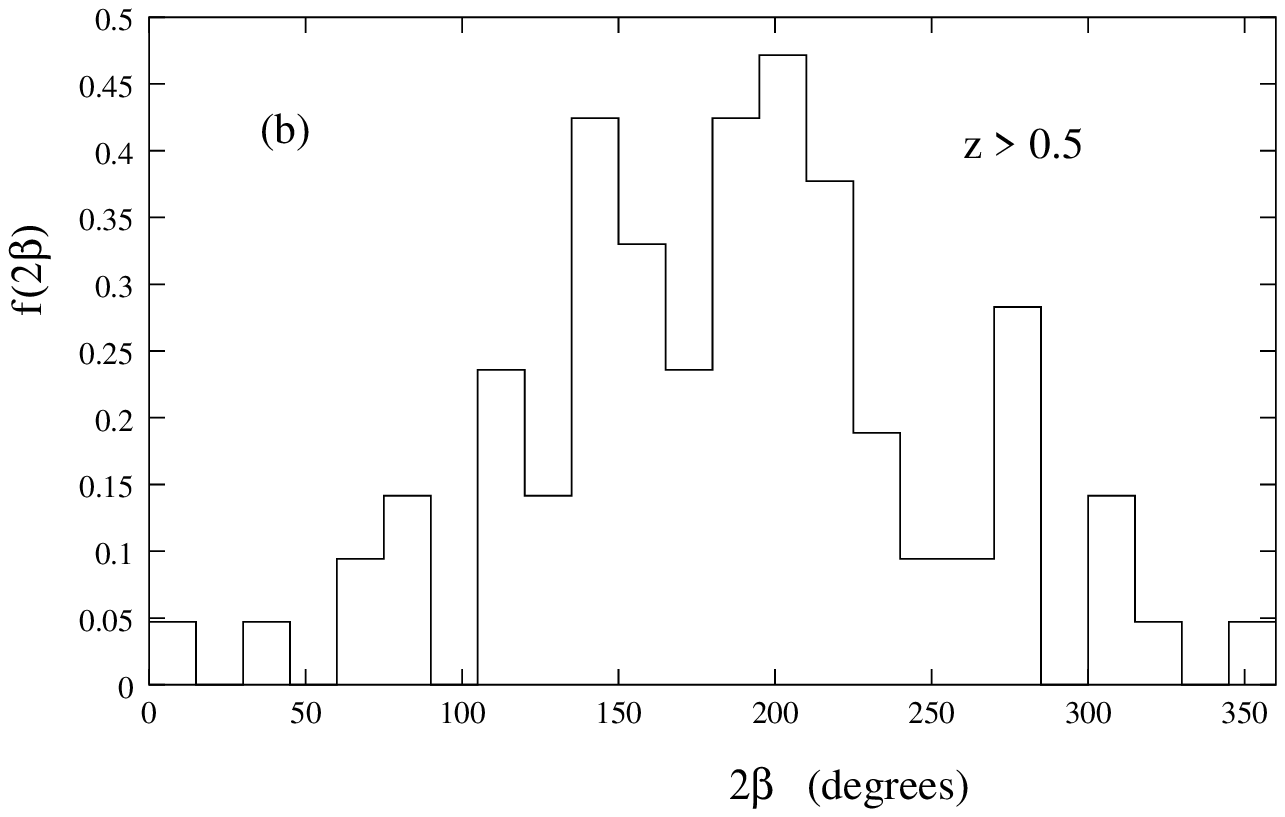,height=6cm}}
} 
\bigskip
\caption{The distribution of the offset angle $\beta$ between the intrinsic
polarization $\chi$ and the orientation axis of the radio galaxy $\psi$ 
for (a) full data containing 231 points and (b) after putting the cut on
redshift, $z>0.5$.
} \label{redshift}
\end{figure}

We next study the 
 redshift, $z$, dependence of distribution.
It turns out that the distribution 
has a strong dependence on the redshift which can be modelled by the
following generalization of the vM distribution,
\begin{equation}
F(\beta,z) = N(\kappa,\kappa_1){\exp \left[- (\kappa + \kappa_1 z)\cos(2\beta)
\right] }
\label{von_z}
\end{equation} 
where $N(\kappa,\kappa_1)$ is a normalization factor. 
Maximizing the likelihood in this case gives $\kappa = 0.29$ and
$\kappa_1= 0.90$. The maximum likelihood increases by 7 units, i.e. 
$L_{II} - L_I = T = 7$, where $L_{I}$ and $L_{II}$ are the maximum
likelihoods for the distributions given in equations (\ref{von}) 
and (\ref{von_z})
respectively. After making the galactic cut the 
corresponding parameters are $\kappa = 0.32$, $\kappa_1= 0.97$
and $T = 5.1$. The statistic $2T$ is distributed as
$\chi^2_1$. The increase in the maximum likelihood is quite large and
implies that the width of the distribution  has significant dependence
on the redshift. This is also apparent from a plot of the distribution
of the angle $2\beta$ shown in Fig. \ref{redshift} for the complete
data set and after imposing the cut on redshift, $z>0.5$. We find that
after imposing the cut the distribution is more strongly peaked at
$\beta=\pi/2$.

In order to search for a correlation of the type implied by 
Eq. \ref{approximate} we propose the following distribution
\begin{equation}
G(\beta,z) = N(\kappa,\kappa_1,\zeta,\xi) {\exp \left[- (\kappa + \kappa_1 z)
\cos(2\beta) + p(z,\zeta,\xi) \sin(2\beta)\right]}
\label{corr}
\end{equation}
where $N(\kappa,\kappa_1,\zeta,\xi)$ is a normalization factor and 
$p(z,\zeta,\xi)$ is given by
\begin{equation}
p(z,\zeta,\xi) = \zeta \left[1-\exp\left(\xi
\left((1+z)^{3/2} -1\right)\right)\right]
\end{equation}
This form of the distribution will lead to the relationship 
(\ref {approximate}) for the polarization
angle $<\beta(z)>$, where $<>$ denotes the mean value, 
 as long as 
deviation of $<\beta>$ from $\pi/2$ is small.

The best fit parameters for the correlated ansatz are
$\kappa=0.27,\ \kappa_1=0.98,\ \zeta= 8.0\times 10^{-14},\ \xi=7.1$
with the statistic $2T = 2(L_{III} - L_{II})= 5.34$ 
for the full data set and $\kappa=0.28,\ \kappa_1=1.1,\ \zeta= 8.8\times 
10^{-8},\ \xi=4.1$
with the statistic $2T = 6.9$ for the case of galactic cut.
Here $L_{III}$ is the maximum
likelihood for the distribution given in equation (\ref{corr}).
The corresponding percentage values (p-values) 
for these two cases, with and without
the galactic cut, are
7\% and 3.5\% respectively. These p-values represent
the probabilities that the correlation seen in data can be
obtained from a statistical fluctuation.  
After making the galactic cut the results correspond to a 
$2\sigma$ effect. 
Hence these results show marginal statistical significance.
Although they show a significant correlation it 
is clear from the best fit parameters that the solution prefers
the extreme values of the parameter $\zeta$ very close to zero.
This implies that the deviation of $\beta$ from
$\pi/2$ is very close to zero
for a wide range of redshifts and deviates from zero only for 
relatively large redshifts. This is clearly seen in Fig. 3 where
we plot the data and our best fit model for the case of galactic cut.
On the y-axis we plot the $\sin 2\beta$ 
in order to deal with quantities invariant under 
angular coordinate transformations.  
The data, $\sin 2\beta$, has been averaged over the redshift interval of 0.25
in order to reduce the noise. The number of data points in each bin,
starting from the smallest redshift,
are 54,19,15,13,10,9,9,11,7,4,3,2,2,1,1 respectively. 

\begin{figure}
\psfig{file=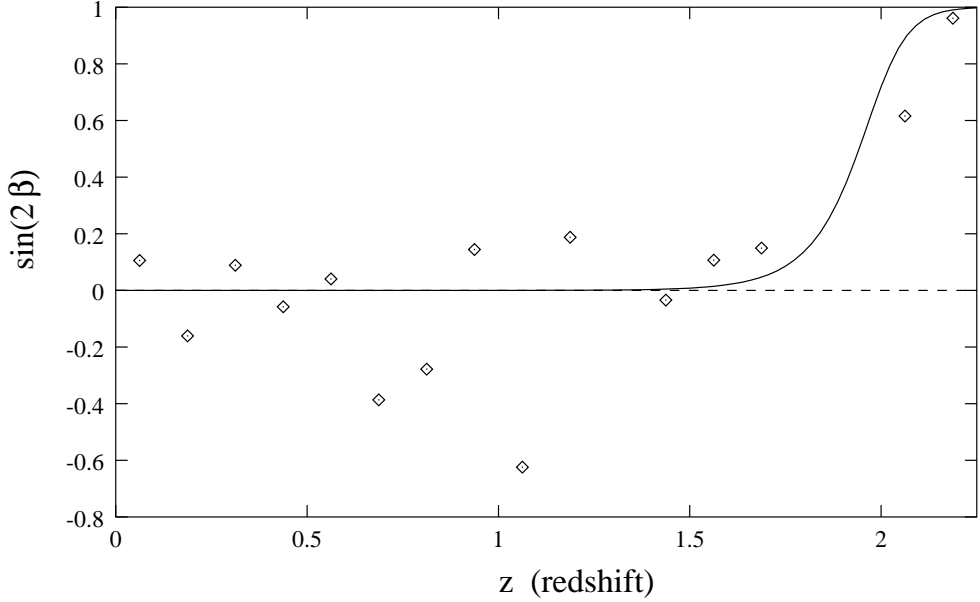}
\bigskip
\caption{The fit to the polarization offset angle $\beta$ for distant radio
galaxies and quasars. The redshift is plotted on the x-axis and the
mean (over a redshift interval of 0.25) of $\sin(2\beta)$ is plotted
on the y-axis. The region within $\pm 30^o$ of the galactic plane
was eliminated from the data.
}
\end{figure}

If we eliminate the two largest redshift points then we do not
see any significant effect. Strictly speaking we were unable to 
find the maxima of the likelihood. The fit was driven to ridicuously
large values of
$\xi>100$ and we did not search beyond these extreme values. The statistic
$2T$ in this case was roughly $1.6$ which has p-values of 45\%
and hence shows no significant correlation. 
In Fig. 4 we show the limits of the parameters $\zeta$ and $\xi$ that
are imposed by the polarization data. The limits are obtained by
excluding all sources which lie within $\pm 30^o$ from the galactic plane.
However the results remain unchanged even if this cut is not
imposed. 

\begin{figure} [t,b] \hbox{\hspace{6em}
\hbox{\psfig{figure=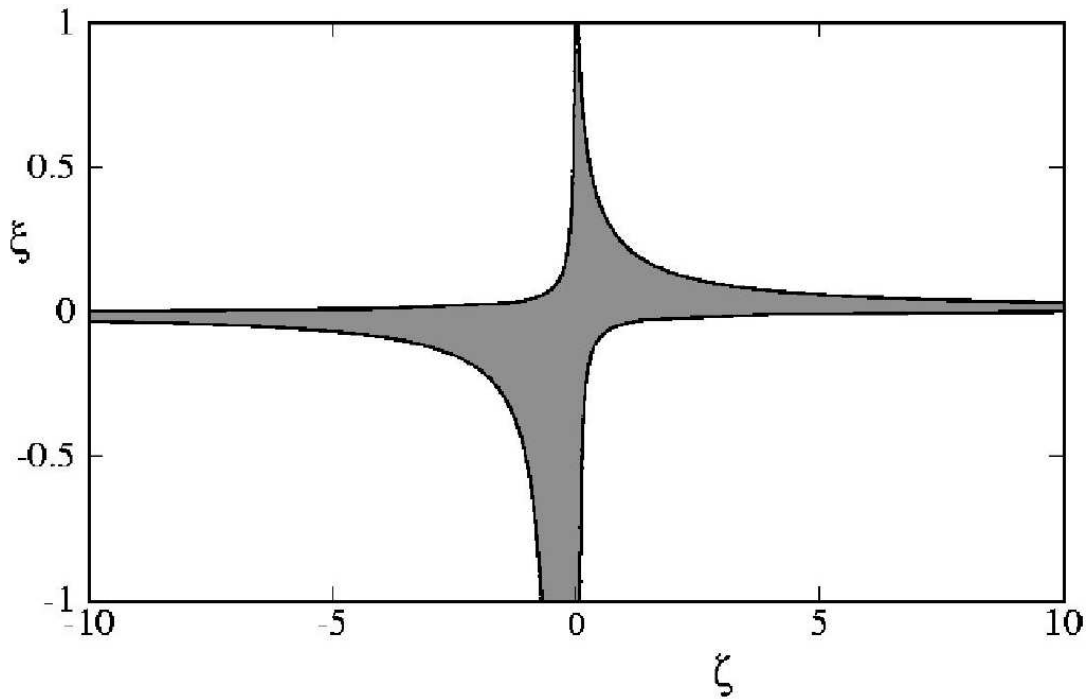,height=10cm}}} \caption{
Limits on the values of the parameters 
$\zeta$ and $\xi$ from the polarization data from distant
quasars and radio galaxies after eliminating all sources
which lie within $\pm 30^o$ from the galactic plane. The results
do not change appreciably if this cut is not imposed. 
The shaded region represents the allowed
range of these parameters. The two largest redshift points were 
deleted for this analysis.
}
\label{limits} \end{figure}

\section{Conclusions}
In this paper, we have studied cosmic birefringence in the framework
of perturbative four-dimensional heterotic string theory. Due to 
coupling between dilaton and axion in this theory, we have
shown that the dilaton field changes the birefringence produced
by the axion field in a nontrivial way. We obtained an approximate
expression for the rotation in polarization predicted by heterotic
string theory within the WKB approximation. We analysed the data for
distant quasars and radio galaxies to determine if the effect is present
in the data. We find a marginally significant effect. However the
correlation is caused primarily by two largest redshift points 
and the effect is lost after these points are eliminated. 
Hence we find that the data does not show significant correlation
but the large redshift points are suggestive that a signal might exist.
This should be thoroughly investigated by accummulating more data
at the largest redshifts.
We have also determined the limits on the string theory parameters
that are imposed by the current polarization data from distant
quasars and radio galaxies.

We, furthermore, like to point out that all these calculations were done 
taking dilaton as a massless scalar. As explained in the introduction,
one of the biggest problem in string theory is to find a mechanism to
remove this massless dilaton from present day physics. 
One expects
that a potential for dilaton will be generated in string theory through
some non-perturbative effects and the dilaton will sit at the minimum of
the potential picking up a mass. As the mecanism of how
precisely this should
happen is still not clear in string theory, we have not dealt with such a
scenario in this paper.

\bigskip
\noindent
{\bf Acknowledgements:} PJ thanks John Ralston for very useful discussions
and for providing very useful insights into the statistical analysis of
this data. 
This research has made use of the NASA/IPAC 
Extragalactic Database (NED) which is
operated by the Jet Propulsion Laboratory, California Institute of Technology, 
under contract with the
National Aeronautics and Space Administration.

\end{document}